\documentclass[aps,prl,twocolumn,amsmath,amssymb,showpacs, superscriptaddress,notitlepage,longbibliography,10pt]{revtex4-1}

\usepackage[colorlinks=true,linkcolor=blue,anchorcolor=red,citecolor=blue, urlcolor=blue]{hyperref}
\DeclareMathAlphabet\mathbfcal{OMS}{cmsy}{b}{n}
\usepackage{bm}
\usepackage{graphicx}
\usepackage{physics}
\usepackage{color}
\usepackage{multirow}
\usepackage{extarrows}
\usepackage{array}
\usepackage{appendix}
\usepackage{amssymb}
\usepackage{physics}
\usepackage{bbding}
\usepackage{xcolor}
\usepackage{tensor}

\usepackage{cases}
\usepackage[normalem]{ulem}

\begin{document}

\title{Quantum Christoffel Nonlinear Magnetization}

\author{Xiao-Bin Qiang}
\email{These authors contributed equally to this work.}
\affiliation{State Key Laboratory of Quantum Functional Materials, Department of Physics, and Guangdong Basic Research Center of Excellence for Quantum Science, Southern University of Science and Technology (SUSTech), Shenzhen 518055, China}

\author{Xiaoxiong Liu}
\email{These authors contributed equally to this work.}
\affiliation{State Key Laboratory of Quantum Functional Materials, Department of Physics, and Guangdong Basic Research Center of Excellence for Quantum Science, Southern University of Science and Technology (SUSTech), Shenzhen 518055, China}

\author{Hai-Zhou Lu}
\email{Corresponding author: luhz@sustech.edu.cn}
\affiliation{State Key Laboratory of Quantum Functional Materials, Department of Physics, and Guangdong Basic Research Center of Excellence for Quantum Science, Southern University of Science and Technology (SUSTech), Shenzhen 518055, China}
\affiliation{Quantum Science Center of Guangdong-Hong Kong-Macao Greater Bay Area (Guangdong), Shenzhen 518045, China}

\author{X. C. Xie}
\affiliation{International Center for Quantum Materials, School of Physics, Peking University, Beijing100871, China}
\affiliation{Interdisciplinary Center for Theoretical Physics and Information
Sciences (ICTPIS), Fudan University, Shanghai 200433, China}
\affiliation{Hefei National Laboratory, Hefei 230088, China}

\date{\today}

\begin{abstract}
The Christoffel symbol is an essential quantity in Einstein's general theory of relativity. We discover that an electric field can induce a nonlinear magnetization in quantum materials, described by a Christoffel symbol defined in the Hilbert space of quantum states (quantum Christoffel symbol). Quite different from the previous scenarios, this orbital magnetization does not need spin-orbit  coupling and inversion symmetry breaking. Through symmetry analysis and first-principles calculations, we identify a number of point groups and 2D material candidates (e.g., BiF$_3$, ZnI$_2$, and Ru$_4$Se$_5$) that host this quantum Christoffel nonlinear magnetization. More importantly, this nonlinear magnetization allows the quantum Christoffel symbol to be probed by optical techniques such as magneto-optical Kerr spectroscopy or transport measurements such as tunneling magneto-resistance. This quantum Christoffel nonlinear magnetization gives a paradigm of how geometry dictates physics.
\end{abstract}
\maketitle

\begin{figure}[htbp]
\centering 
\includegraphics[width=0.48\textwidth]{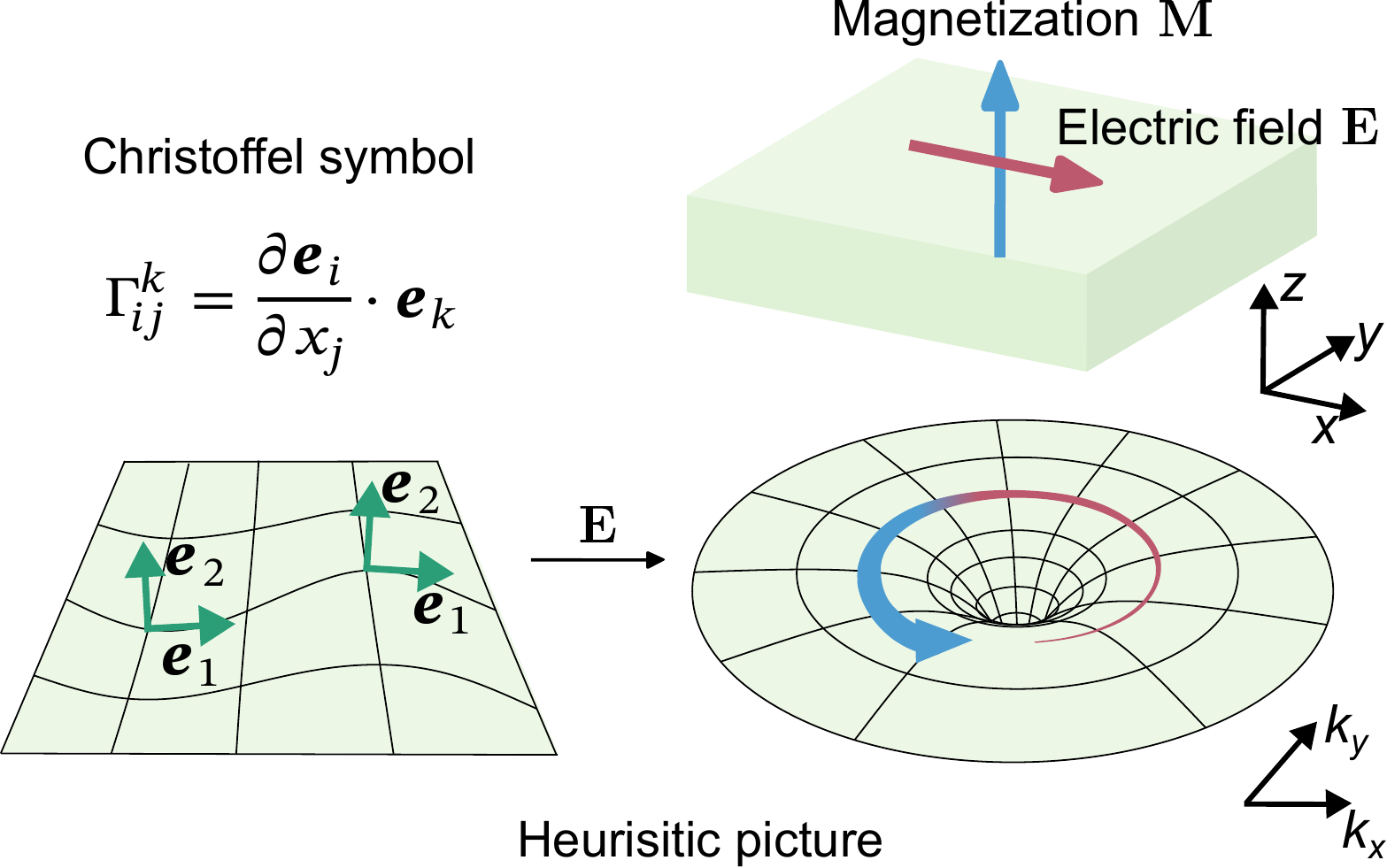}
\caption{In Einstein's general relativity, the Christoffel symbol $\Gamma_{ij}^k$ describe how basis vectors $\boldsymbol{e}_i$ transform across the coordinates $\boldsymbol{x}$ of the curved spacetime. We find a nonlinear magnetization $\mathbf{M}$ in response to an electric field $\mathbf{E}$ can be described by a quantum Christoffel symbol $\boldsymbol{\Gamma}$ in momentum space ($k_x$-$k_y$), in terms of $\mathbf{M}\propto \mathbf{E} \mathbf{E} \int \boldsymbol{\Gamma}$. A heuristic picture is as follows. One role of the electric field $\mathbf{E}$ is to induce a curved quantum space, described by $\boldsymbol{\Gamma}$.
The other role of $\mathbf{E}$ is to drive a ``loop current" of electron orbital motion in the curved quantum space, leading to the magnetization $\mathbf{M}$ that is quadratic in $\mathbf{E}$.}
\label{Fig: demo}
\end{figure}

\textit{\textcolor{blue}{Introduction.}--} The Christoffel symbol defines how the basis vectors transform in the curved spacetime of Einstein's general theory of relativity (Fig.~\ref{Fig: demo}), giving the essential geometric description of gravity~\cite{KPThorne73book,Wald10book}. The geometry is also tied to a number of nontrivial phenomena in quantum materials~\cite{Provost80cmp,Berry84rspa,AA90prl,XiaoD10rmp,Resta11epjb,Torma23prl,Lu24nsr,YanBH25rppp}. In particular, Berry curvature and quantum metric, as the imaginary and real parts of the quantum geometric tensor, have been found playing critical roles in the quantum anomalous Hall effect~\cite{Haldane88prl,MacDonald10rmp,Lu10prb,YuR10science,XueQK13science,ChangCZ23rmp}, nonlinear transport~\cite{GaoY14prl,FuL15prl,Lu18prl,Lu21nrp,GaoY21prl,YangSY21prl,XuSY19nature,KangKF19nm,Tiwari21nc,Kumar21nn,MaT22nc,KTLaw23nsr,XuSY23science,GaoWB23nature,WangJ24prb,Lu25as,YangSY25prl}, flat-band superconductivity~\cite{Peotta15nc,Julku16prl,Torma17prb,Torma18prb,Torma22prb}, and fractional Chern insulators~\cite{Roy12prb,Roy14prb,Jackson15nc}. On the other hand, the modern theory of magnetization reveals that the 
essential role of the rotation of wave packets~\cite{Niu05prl,Resta05prl,Resta06prb,Niu07prl,Krister19jpcs,DongL20prl,Dimi24apx,Lu26apl}. The seminal works have demonstrated that the orbital contribution can give rise to linear magnetization in crystalline solids without spin-orbit coupling and in the absence of inversion symmetry~\cite{Yoda15sr,Moore16prl,Pesin17prb,Murakami20prb, Falko20prl,Falko23nn,Johansson24jpcm}. These advances have revealed that orbital magnetization is intimately related to the Berry curvature, thereby providing a geometric perspective on the microscopic origin of magnetization. However, in inversion symmetry broken systems, it is challenging to intrinsically distinguish the spin and orbital contributions in experiments because of spin-orbit coupling.

In this Letter, we discover a nonlinear orbital magnetization in quantum materials, described by a quantum version of the Christoffel symbol of the first kind.  As shown in Fig. \ref{Fig: demo}, this orbital magnetization $\mathbf{M}$ is a quadratic nonlinear response to an electric field $\mathbf{E}$. We find that the quantum Christoffel symbol plays an indispensable role in the quadratic relation between $\mathbf{M}$ and $\mathbf{E}$, even dictates for Dirac fermions with particle-hole symmetry [see Eqs.~\eqref{Eq: alpha_CS} and \eqref{Eq: Christoffel}]. 
Quite different from the previous scenarios~\cite{Yoda15sr,Moore16prl,Pesin17prb,Murakami20prb,Edestein90ssc,Kato04prl,Manchon19rmp,YangSY22prl,YangSY23prl}, this quantum Christoffel nonlinear magnetization can survive in absence of the spin-orbit coupling in nonmagnetic materials, and does not require inversion symmetry breaking. Through symmetry analysis, we find that the quantum Christoffel nonlinear magnetization can be allowed by many point-group symmetries and hosted by a number of 2D candidate materials. As an example, we use first-principles calculations to evaluate the nonlinear magnetization coefficient for the material BiF$_3$ and find that the quantum Christoffel nonlinear magnetization is the main contribution in it. Moreover, our calculations indicate that this quantum Christoffel nonlinear magnetization gives rise to a measurable Kerr signal and can be probed via magneto-optical Kerr spectroscopy~\cite{Sugano13book,Kuch15book}. It may also explain the observed anomalous second-order tunneling magneto-resistance~\cite{LiaoZM24prb}.

\textit{\textcolor{blue}{Quantum Christoffel nonlinear magnetization.}--} To drive the nonlinear magnetization coefficient $\alpha_{ijk}$ defined as $M_i=\alpha_{ijk}E_j E_k$ ($i,j,k\in \{x,y,z\}$), we calculate the quadratic nonlinear orbital magnetization $\mathbf{M}$ induced by an electric field $\mathbf{E}$ (Details in Secs.~SI-SIII of Supplemental Material~\cite{Supp}). Suppose the quantum state $\ket{\nu\mathbf k}$ in a crystal is described 
by the energy band index $\nu $ and wave vector $\mathbf{k}$, the orbital magnetization are summation of the orbital magnetic moment $\tilde{\mathbf{m}}_\nu(\mathbf{k})$ weighted by the non-equilibrium distribution function $f_\nu(\mathbf{k})$, as 
\begin{equation}\label{Eq: M_def}
\mathbf{M}=\frac{1}{\mathcal{V}}\sum_{\nu,\mathbf{k}}\tilde{\mathbf{m}}_\nu(\mathbf{k}) f_\nu(\mathbf{k}),
\end{equation}
where $\mathcal{V}$ is volume. Both $\tilde{\mathbf{m}}_\nu(\mathbf{k})$ and  $f_\nu(\mathbf{k})$ are modified to include nonequilibrium components induced by the driving electric field $\mathbf{E}$.

In our treatment, $f_\nu(\mathbf{k})$ is derived from the Boltzmann equation~\cite{Haug08book,Lu19nc,Lu23prb} and $\tilde{\mathbf{m}}_\nu(\mathbf{k})$ is evaluated by applying the wave packet dynamics up to the leading order of $\mathbf{E}$~\cite{GaoY14prl,GaoY15prb,GaoY19fop,XiaoC21prb}. Unless necessary, we suppress $\mathbf{k}$ for simplicity.

We find that, for 2D Dirac fermions with particle-hole symmetry, that is, the conduction ($c$) and valence ($v$) band energies have the relation $\varepsilon_c(\mathbf{k})=-\varepsilon_v(\mathbf{k})$, the nonlinear magnetization coefficient
\begin{equation}\label{Eq: alpha_CS}
\alpha_{ijk} =-\frac{e^3}{2\hbar}\frac{\tau }{\mathcal{V}}\sum_{\nu,\mathbf{k}} \epsilon_{ilr} \Gamma_\nu^{rlj}v_\nu^k  f_0',
\end{equation}
where $i,j,k,l,r\in \{x,y,z\}$ with Einstein summation convention implied over repeated indices, $-e$ is the electron charge, $\tau$ is the relaxation time, $\epsilon_{ilr}$ is the Levi-Civita anti-symmetric tensor, $\boldsymbol{v}_\nu=\nabla_\mathbf{k} \varepsilon_\nu(\mathbf{k})/\hbar$ is the intraband velocity,  
$\varepsilon_\nu(\mathbf{k})$  is the eigen energy for a state with wave vector $\mathbf{k}$ on band $\nu$, and the Fermi-Dirac function $f_0=[e^{(\varepsilon_\nu-\varepsilon_F)/k_BT}+1]^{-1}$, with Fermi energy $\varepsilon_F$, Boltzmann constant $k_B$, and temperature $T$. Here, $f_0'$ is the derivative of $f_0$ with respect to $\varepsilon_\nu$, so $\alpha_{ijk}$ is primarily contributed by the states near the Fermi surface. More importantly, $\alpha_{ijk}$ explicitly contains a quantum Christoffel symbol
\begin{equation}\label{Eq: Christoffel}
\Gamma_\nu^{rlj}=\frac{1}{2}(\partial_j g_\nu^{lr}+\partial_l g_\nu^{rj}-\partial_r g_\nu^{lj}),
\end{equation}
where $\partial_j$ represents $\partial/(\partial k_j)$, and $\mathbf{g}_\nu = \text{Re} \sum_{\mu \neq \nu} \mathbfcal{A}_{\nu\mu } \mathbfcal{A}_{\mu \nu}$ is the quantum metric tensor~\cite{Provost80cmp,Resta11epjb,Torma23prl,Lu24nsr,YanBH25rppp} defined by the interband Berry connection
$\mathbfcal{A}_{\mu \nu}=i\langle \mu |\nabla_{\mathbf k}|\nu\rangle$~\cite{Berry84rspa}. The Christoffel symbol originally characterizes the curved spacetime in Einstein's general relativity through the geodesic equation~\cite{KPThorne73book,Wald10book} (\hyperlink{Appendix A}{Appendix A})
\begin{equation}
\ddot{x}_k + \Gamma^k_{ij} \dot{x}_i \dot{x}_j = 0,
\end{equation}
where $\ddot{x}_k$ and $\dot{x}_{i,j}$ are the derivatives of the coordinates $x_i$ with respect to proper time. In our context, the quantum Christoffel symbol is defined in the Hilbert space of quantum states, and the electric field and magnetization have the analogies
\begin{eqnarray}
E_i \sim \dot{x}_i, \ \ \ M_k \sim \ddot{x}_k.
\end{eqnarray}
Recently, the quantum Christoffel symbol has also been introduced to describe momentum-space gravity~\cite{Smith22prr,Mehraeen25prl,FuL26prb}. 

We now clarify how the quantum Christoffel symbol mediates the quadratic relationship between the orbital magnetization and the electric field. The quantum metric tensor $ \mathbf{g}_\nu$ measures the distance between quantum states, thus dictates the trajectory of a moving electron. We now give a hand-waving argument on how the quantum Christoffel symbol gives rise to magnetization. Eq.~\eqref{Eq: Christoffel} implies that the Einstein summation of the Levi-Civita tensor $\epsilon_{ilr}$ with the quantum Christoffel symbol $\Gamma_\nu^{rlj}$ in Eq.~\eqref{Eq: alpha_CS} can be interpreted as the curl of the quantum metric tensor, i.e., $\varepsilon_{ilr}\Gamma_\nu^{rlj} = (\nabla_\mathbf{k} \times \mathbf{g}_\nu)^{ij}$. This induce an effective loop current under an applied electric field, giving rise to an orbital magnetization. Moreover, one of the roles of the electric field is to induce the curl of quantum metric tensor, the other is to drive the orbital rotation in the curved space of quantum states, so this magnetization is quadratic in electric field.

\textit{\textcolor{blue}{Particle-hole symmetry and dominance of quantum Christoffel symbol.}--}
Why particle-hole symmetry guarantees the dominance of the quantum Christoffel symbol in the nonlinear orbital magnetization of Dirac fermions can be understood as follows. Dirac fermions can generically describe a number of semiconductors or insulators with or without topological properties~\cite{Shen17book}. The Hamiltonian of 2D Dirac fermions can generally be written as $\mathcal{H} = d_0 + \mathbf{d} \cdot \boldsymbol{\sigma}$, where $d_0$ and $\mathbf{d} = (d_x, d_y, d_z)$ are functions of the wave vector $\mathbf{k}$, and $\boldsymbol{\sigma} = (\sigma_x, \sigma_y, \sigma_z)$ is the vector of Pauli matrices. In a general case that particle-hole symmetry is not necessary, the nonlinear magnetization coefficient $\alpha_{ijk}$ is found to take the form
\begin{equation}\label{Eq: alpha_full}
\alpha_{ijk} =\frac{\tau e^2}{\mathcal{V}}\sum_{\nu,\mathbf{k}} F_\nu^{ij} v_\nu^k f_0',
\end{equation}
where the magnetoelectric tensor $\mathbf{F}_\nu$ 
is a band geometric quantity, given by
\begin{equation}\label{Eq: MET}
F_\nu^{ij}=-2\text{Re} \sum_{\mu \neq \nu} \frac{m_{\nu \mu}^i \mathcal{A}_{\mu \nu}^j}{\varepsilon_\nu-\varepsilon_\mu}
-\frac{e}{2\hbar}\epsilon_{ilr}\Gamma_\nu^{rlj}.
\end{equation}
Here, $\mathbf{m}_{\mu\nu}=e\sum_{\rho\neq \nu}(\boldsymbol{v}_{\mu \rho}+\delta_{\rho\mu }\boldsymbol{v}_\nu)\times \mathbfcal{A}_{\rho\nu}/2$ is the interband orbital magnetic moment~\cite{GaoY14prl,YangSY24prl}, where $\boldsymbol{v}_{\mu \nu}=\langle \mu| \nabla_\mathbf{k} \mathcal{H} | \nu \rangle/\hbar$ is the interband velocity. For the Hamiltonian of Dirac fermions, we find that (Sec.~IV of~\cite{Supp}), 
\begin{equation}
F_{\pm}^{ij}=\mp\frac{e}{\hbar} \epsilon_{ikl}\frac{\partial_k d_0}{d}
g_{\pm}^{l j}-\frac{e}{2\hbar}\epsilon_{ilr}\Gamma_\pm^{rlj},
\end{equation}
where $d = |\mathbf{d}|$, and $+/-$ correspond to the conduction and valence bands, respectively. In the presence of particle-hole symmetry, $d_0 = 0$, then the first term of $F_{\pm}^{ij}$ vanishes and only the quantum Christoffel symbol contributes to the nonlinear orbital magnetization.

\begin{table}[htbp]
\caption{Symmetry constraints of the nonlinear magnetization coefficient $\alpha_{ijk}$ for all point-group symmetry operations. $\mathcal{C}_i^n$ denotes the $n$-fold rotation ($n=2,3,4,6$) with respect to the Cartesian axis along the $i$-direction, and $\mathcal{M}_i$ represents a mirror with respect to the plane normal to the $i$-direction. Symbols \Checkmark and \XSolidBrush mean symmetry-allowed and symmetry-forbidden, respectively. The improper rotations (not listed here) can be derived through $\mathcal{S}_i^n = \mathcal{C}_i^n\mathcal{M}_i$.} \label{Tab: Symm_pg}  
\centering
\setlength{\tabcolsep}{3.7pt} 
\renewcommand{\arraystretch}{1.5} 
\begin{tabular}{ccccccccccc}
\hline\hline
\quad & $\mathcal{C}_z^2$ & $\mathcal{C}_z^3$ & $\mathcal{C}_z^{4,6}$ & $\mathcal{C}_x^{2,4,6}$ & $\mathcal{C}_y^{2,4,6}$ & $\mathcal{C}_x^3$ & $\mathcal{C}_y^3$ & $\mathcal{M}_z$ & $\mathcal{M}_x$ & $\mathcal{M}_y$\\
\hline
$\alpha_{xxx}$ & \XSolidBrush & \Checkmark & \XSolidBrush  & \Checkmark &\XSolidBrush & \Checkmark & \Checkmark & \XSolidBrush & \Checkmark & \XSolidBrush\\
$\alpha_{zzz}$ & \Checkmark  & \Checkmark  & \Checkmark  & \XSolidBrush & \XSolidBrush & \Checkmark & \Checkmark & \Checkmark & \XSolidBrush & \XSolidBrush\\
$\alpha_{xyy}$ & \XSolidBrush & \Checkmark & \XSolidBrush & \Checkmark & \XSolidBrush & \Checkmark & \XSolidBrush & \XSolidBrush & \Checkmark & \XSolidBrush\\
$\alpha_{zxx}$ & \Checkmark & \Checkmark & \Checkmark & \XSolidBrush & \XSolidBrush & \XSolidBrush & \Checkmark  & \Checkmark & \XSolidBrush & \XSolidBrush\\
\hline\hline
\end{tabular}
\end{table}

\textit{\textcolor{blue}{Symmetry analysis and 2D material candidates.}--} Nonzero components of $\alpha_{ijk}$ can be determined for all crystallographic point group operations, following the general tensor transformation rule. Under an arbitrary symmetry operation represented by the matrix $\mathcal{O}$, the rank-3 tensor $\alpha_{ijk}$ transforms according to $\alpha_{i'j'k'} = \text{Det}(\mathcal{O})\mathcal{O}_{i'i}\mathcal{O}_{j'j} \mathcal{O}_{k'k} \alpha_{ijk}$, where the determinant factor $\text{Det}(\mathcal{O})$ accounts for the axial nature of $\alpha_{ijk}$. The symmetry constraints on $\alpha_{ijk}$ for all crystallographic point group operations are summarized in Table~\ref{Tab: Symm_pg}. Specifically, we consider a convenient scenario, $\alpha_{zxx}$, i.e, the quantum Christoffel nonlinear magnetization polarized along the $z$ direction induced by an $x$-direction electric field. This response is particularly relevant in experimental settings. For example, under the mirror reflection $\mathcal{M}_x$, the $z$-component of magnetization $M_z$ changes sign ($M_z \rightarrow -M_z$), while $E_x^2$ remains invariant. As a result, the mirror symmetry $\mathcal{M}_x$ forbids a nonzero $\alpha_{zxx}$. Furthermore, according to the last row of Table~\ref{Tab: Symm_pg}, $\alpha_{zxx}$ is also forbidden by the $\mathcal{C}_{x,y}^2$ and $\mathcal{M}_{x,y}$ symmetries. Therefore, these symmetries must be broken to allow a finite quantum Christoffel nonlinear magnetization, as further illustrated by the model study in \hyperlink{Appendix B}{Appendix B}.

The symmetry analysis presented in Table~\ref{Tab: Symm_pg} indicates that the quantum Christoffel nonlinear magnetization polarized along the $z$-direction is allowed only in materials classified in the point groups 1, $\overline{1}$, 4, $\overline{4}$, 4/m, 3, $\overline{3}$, 6, $\overline{6}$, and 6/m. Based on these symmetry criteria, we utilize the 2DMatPedia database~\cite{2DMat} to identify a set of 2D materials that support the quantum nonlinear Christoffel magnetization, as listed in Table~\ref{Tab: candidates}. Among them, the candidates with point group symmetries 4/m, $\overline{3}$, and 6/m simultaneously exhibit inversion and time-reversal symmetries, while the others retain only time-reversal symmetry.

\begin{table}[htbp]
\caption{2D material candidates that could host $\alpha_{zxx}$, the second-order quantum Christoffel nonlinear magnetization polarized along the $z$ direction induced by an $x$-direction electric field, based on a search in the open computational database 2DMatPedia~\cite{2DMat} of 2D materials as well as verification by the first-principles calculations. The point groups 1 and $\overline 1$ are discarded. More 2D material candidates respect the symmetry can be found in Table~S2 of~\cite{Supp}.} 
\label{Tab: candidates}  
\centering\setlength{\tabcolsep}{3.7pt} 
\renewcommand{\arraystretch}{1.5} 
\begin{tabular}{cc}
\hline\hline
Point group & Candidates\\
\hline
$4$ & TiPb$_9$O$_{11}$ \\
$\overline 4$ &  Sn$_3$(HO$_2$)$_2$, AgBrO$_4$ \\
$4/m$ &  Ru$_4$Se$_5$, VPO$_5$\\
$3$ &   LiSnCl$_3$, CuBi(PSe$_3$)$_2$ \\
$\overline 3$ & BiF$_3$, ZnI$_2$\\
$6$ &  -- \\
$\overline 6$ & AgNO$_3$, Ge$_3$Sb$_2$O$_9$ \\
$6/m$ & -- \\
\hline\hline
\end{tabular}
\end{table}

\begin{figure*}[htbp]
\centering 
\includegraphics[width=0.99\textwidth]{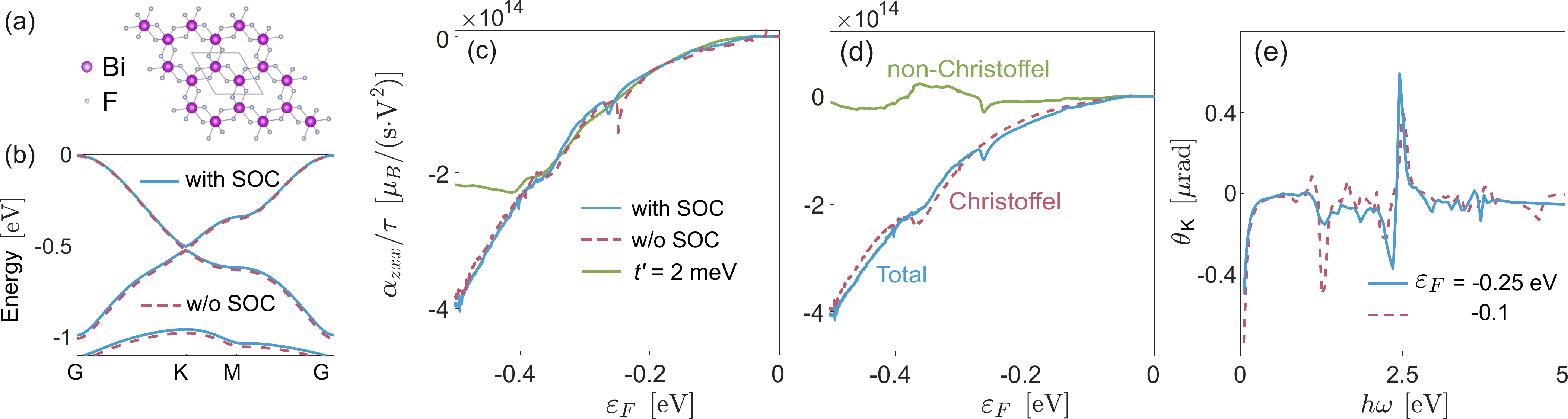}
\caption{(a) Lattice structure of the 2D BiF$_3$ grown along the [001] crystallographic direction. (b) Band structure of the 2D BiF$_3$ along the high symmetry line with and without spin-orbit coupling (SOC). (c) First-principles calculation results of the nonlinear magnetization coefficient $\alpha_{zxx}$ as a function of the Fermi energy $\varepsilon_F$, by using Eq.~(\ref{Eq: alpha_full}) (with and without SOC) and also with the Haldane-like perturbation $\mathcal{H}'$ Eq.~(\ref{Eq: Hp}) ($t'$ = 2 meV), respectively. In the calculation with $\mathcal{H}'$, the hopping parameter $t'$ = 2 meV, relaxation time $\tau$ = 100 ps, and additional electric field $\mathbf{E}$ = 0.02 V/$\mu$m. (d) First-principles calculation results of $\alpha_{zxx}$ as a function of the Fermi energy $\varepsilon_F$, by using Eq.~(\ref{Eq: alpha_CS}) (Christoffel) and Eq.~(\ref{Eq: alpha_full}) (Total), respectively. The non-Christoffel contribution is much smaller than the contribution described by the quantum Christoffel symbol. (e) Kerr rotation angle $\theta_\text{K}$ as a function of photon energy $\hbar \omega$, at Fermi level -0.25 eV and -0.10 eV, respectively. The Kerr angle measures how much a linearly polarized light is rotated after being reflected from the sample and is a direct probe of magnetization.}
\label{Fig: DFT}
\end{figure*}

\textit{\textcolor{blue}{BiF$_3$ and experimental implementation.}--}To verify the result of the symmetry analysis, we perform first-principles calculations for one of the candidates BiF$_3$, as an example (technical details in \hyperlink{Appendix C}{Appendix C}). It has a trigonal lattice (space group P$\overline{3}$), as shown in Fig.~\ref{Fig: DFT}(a). Although BiF$_3$ contains the heavy element Bi with strong spin-orbit coupling, the electronic states near the Fermi level are dominated by the $p$-orbitals of the F atoms. As a result, the effective spin-orbit coupling near the Fermi level is minimal, as shown by the band structure in Fig.~\ref{Fig: DFT}(b). By using Eq.~(\ref{Eq: alpha_full}) and the Wannier interpolation method, we calculate the nonlinear magnetization coefficient $\alpha_{zxx}$ as a function of the Fermi energy $\varepsilon_F$, in the presence and absence of spin-orbit coupling. The two cases yield nearly identical results, as shown in Fig.~\ref{Fig: DFT}(c), indicating that the nonlinear magnetization is mainly from orbitals and can occur in systems with weak spin-orbit coupling. The spin contribution~\cite{YangSY23prl} to the nonlinear magnetization is found to be two orders of magnitude smaller than the orbital contribution in BiF$_3$ (Sec. SVI of~\cite{Supp}). More importantly, by comparing Eqs.~(\ref{Eq: alpha_CS}) and (\ref{Eq: alpha_full}),  Fig.~\ref{Fig: DFT}(d) shows that $\alpha_{zxx}$ can be contributed almost entirely by the quantum Christoffel symbol in BiF$_3$.

Moreover, due to the $\mathcal{C}_z^3$ symmetry of BiF$_3$, the magnetization coefficient exhibits the properties $\alpha_{zxx} = \alpha_{zyy}$ and $\alpha_{zxy} = -\alpha_{zyx}$. This symmetry endows an isotropic characteristic,
\begin{equation}\label{Eq: isotropic}
M_z=\alpha_{zxx}|\mathbf{E}|^2,
\end{equation}
which means the magnitude of the induced $z$-polarized nonlinear magnetization remains invariant with respect to the direction of the in-plane electric field. Remarkably, due to the $\mathcal{C}_z^{n\geq 3}$ symmetry, this isotropic behavior emerges universally among the candidates listed in Table~\ref{Tab: candidates}, which is particularly advantageous for experimental implementation.

The quantum Christoffel nonlinear magnetization can be detected by using the magneto-optical Kerr effect, wherein the polarization state of light reflected from a material is altered by magnetization. 
The angle between the major axis of the incident linearly polarized light and the major axis of the reflected elliptically polarized light is the Kerr rotation angle $\theta_\text{K}$. It can be simulated from the first-principles calculations by using the formula~\cite{Feng_DFT_Kerr_2015, Guo_Kerr_1994,Guo_Kerr_1995}
\begin{equation}\label{Eq: theta_K}
\theta_\text{K} =  {\rm Re} \frac{\epsilon_{xy}}{(\epsilon_0-\epsilon_{xx}) \sqrt{\epsilon_{xx}/(d\epsilon_0)}} , 
\end{equation}
where $\epsilon_0$ is the vacuum dielectric constant, $d$ is the sample thickness, and $\epsilon_{xy}$ and $\epsilon_{xx}$ are dielectric functions. In the presence of an external electric field $\mathbf{E}$,  $\epsilon_{xy}$ can be expanded up to the second order of $\mathbf{E}$, $\epsilon_{xy}=\epsilon_{xy}^{(0)}+\epsilon_{xy}^{(1)}(\propto\mathbf{E})+\epsilon_{xy}^{(2)}(\propto\mathbf{E}^2)+\cdots$, where the equilibrium dielectric function $\epsilon_{xy}^{(0)}$ and first-order nonequilibrium dielectric function $\epsilon_{xy}^{(1)}$ are forbidden by time-reversal and inversion symmetries, respectively~\cite{Tsirkin18prb}. Therefore, the only nonzero term is $\epsilon_{xy}^{(2)}$. In practice, the dielectric functions are determined by the optical conductivities as $\epsilon_{xx}=\epsilon_0+i\sigma_{xx}(\omega)/\omega$ and $\epsilon_{xy}^{(2)}=i\sigma_{xy}(\omega)/\omega$, where $\omega$ is the light frequency ~\cite{Sugano13book,Kuch15book}. The optical conductivity $\sigma_{ij}(\omega)$ can be calculated by using the Kubo-Greenwood formula~\cite{Kubo:1957,Greenwood:1958}
\begin{equation}\label{Eq: opt-cond}
\sigma_{ij}(\omega)=\frac{e^2 \hbar}{i \mathcal{V}}  \sum_{\nu, \mu, \mathbf{k}} \frac{f_0\left(\varepsilon_\nu\right)-f_0\left(\varepsilon_\mu\right)}{\varepsilon_\mu-\varepsilon_\nu} \frac{\left\langle \nu\right| v_i \left|\mu \right\rangle\left\langle \mu \right| v_j \left| \nu \right\rangle} {\left(\varepsilon_\mu-\varepsilon_\nu\right)-\hbar \omega-i \eta}~,
\end{equation} 
where $v_i=\nabla_{k_i}\mathcal{H}/\hbar$ is the velocity operator along the $i$ direction, $\hbar\omega$ is the photon energy, and $\eta$ is an infinitesimal. 

To evaluate the Kerr angle using first-principles calculations, we introduce a Haldane-like perturbation $\mathcal{H}'$ to simulate the effect of magnetization
\begin{equation}\label{Eq: Hp}
\mathcal{H}'=\sum_{\langle\langle i,j\rangle\rangle}  \sum_{a,b}  t' e^{i \theta_{ij,ab}} c_{i,a}^\dagger c_{j,b},
\end{equation}
where $\langle\langle i,j\rangle\rangle$ denotes the second-nearest sites of Bi atoms, $a,b \in \{p_x,p_y\}$ with $a\neq b$, and only the hoppings between $p_x, p_y$ orbitals of Bi atoms are considered. To maintain the $\mathcal{C}_z^3$ symmetry, the hopping phase factor $\theta_{ij, p_xp_y}$  is chosen with respect to the hopping direction, taking values (0, $\pm$ 2$\pi/3$, $\pm$ 4$\pi/3$). Crucially, a 2$\pi$/3 difference between $\theta_{ij, p_xp_y}$ and $\theta_{ij, p_yp_x}$ is needed to break time-reversal symmetry. 

As shown in Fig.~\ref{Fig: DFT}(c), the behavior of $\alpha_{zxx}$ near the Fermi energy can be well described by the Wannier Hamiltonian with the  perturbation  $\mathcal{H}'$ for a  hopping magnitude of $t'$ = 2 meV. Moreover, $\alpha_{zxx}$ for the finite hole doping can induce a nonlinear orbital magnetic moment per unit cell of the order of $10^{-4} \mu_B$, if we assume that the BiF$_3$ sample is in an in-plane electric field of the order of 0.02 V/$\mu$m and the relaxation time $\tau$ = 100 ps. This ensures that the magnetization can be detected by the magneto-optical Kerr spectroscopy.

Fig.~\ref{Fig: DFT}(e) displays the simulated Kerr rotation angle $\theta_{\text{K}}$ as a function of photon energy $\hbar\omega$, calculated by using the perturbed Wannier Hamiltonian for the Fermi energies at $-0.25$ and $-0.10$ eV. The predicted Kerr rotation amplitudes are on the order of $\mu\mathrm{rad}$, which is well above the typical experimental detection limit (on the order of $n\mathrm{rad}$~\cite{Choi23nature}). This estimation is based on a relaxation time of $\tau = 100$ ps, and even shorter relaxation times down to the picosecond range can still yield detectable Kerr signals under suitable conditions. In addition, our analysis identifies the 2–3 eV photon energy range as optimal, exhibiting a pronounced Kerr angle $\theta_\text{K}$ across a broad doping window. Here, the Kerr signal of the nonlinear magnetization in time-reversal-symmetric systems can be distinguished from those of the conventional magneto-optical effects as they vanish identically in the presence of time-reversal symmetry. In particular, the inverse Faraday effect requires circularly polarized light and is inherently tied to mechanisms that break time-reversal symmetry~\cite{Stefan11prb,Kruglyak12prb,Oppeneer14prb,Ulrich21prb}. While linearly polarized light can be converted into circular polarization, this conversion unavoidably relies on time-reversal-symmetry breaking. Thus, the nonlinear magnetization uncovered here is fundamentally distinct from previously known magneto-optical effects and constitutes a genuinely new mechanism.

In the above, we have focused on nonmagnetic materials, thereby eliminating all time-reversal-odd contributions (see Secs. SIII and SV of~\cite{Supp}). Moreover, as a second-order response, $\alpha_{ijk}$ can remain nonzero even in centrosymmetric materials and is fundamentally independent of spin-orbit coupling. This opens new avenues for realizing nonlinear orbitronic phenomena~\cite{ZhangSC05prl,Inoue08prb,Yuriy21epl,Choi23nature,WangP24aem,Jo24npj,WuXS24prl,Dimi24prl,Dimi25prl} in light-element materials and high-symmetry systems that were previously considered unsuitable for spin-orbit-driven effects.

\begin{acknowledgments}
We thank Xuan Qian, Kaiyou Wang, Yang Ji, Rui Chen, Chunming Wang, Zongzheng Du, Xiangang Wan, and Xiaoqun Wang for the insightful discussions. This work is supported by the National Key R$\&$D Program of China (2022YFA1403700), the National Natural Science Foundation of China (12525401 and 12350402), Quantum Science and Technology—National Science and Technology Major Project (Grant No. 2021ZD0302400), Guangdong Basic and Applied Basic Research Foundation (2023B0303000011), Guangdong Provincial Quantum Science Strategic Initiative (GDZX2201001 and GDZX2401001), the Science, Technology and Innovation Commission of Shenzhen Municipality (ZDSYS20190902092905285),
High-level Special Funds (G03050K004), the New Cornerstone Science Foundation through the XPLORER PRIZE, and the Center for Computational Science and Engineering of SUSTech.
\end{acknowledgments}

\bibliographystyle{apsrev4-1-etal-title_10authors}
\bibliography{ref}

\appendix 

\renewcommand{\theequation}{A\arabic{equation}}
\setcounter{equation}{0}

\section{{\large End Matter}} 

\hypertarget{Appendix A}{\textit{\textcolor{blue}{Appendix A: Tutorial on Christoffel symbol in Einstein's general relativity.}--}} 
The motion of a free particle in curved spacetime of general theory relativity is governed by the geodesic equation
\begin{equation}\label{Eq: Geodesic}
\ddot{x}_k + \Gamma^k_{ij} \dot{x}_i \dot{x}_j = 0,
\end{equation}
which generalizes Newton's second law to a curved spacetime, where $\ddot{x}_k$ and $\dot{x}_{i,j}$ denote the derivatives of $x_i$ with respect to proper time. The geometric properties of spacetime are encoded in the affine connection
\begin{equation}
\Gamma_{ij}^k=\frac{\partial\boldsymbol{e}_i}{\partial x_j}\cdot\boldsymbol{e}_k,
\end{equation}
which characterizes the variation of the basis vectors $\boldsymbol{e}_i$'s throughout a given coordinate system (Fig.~\ref{Fig: demo}), 
where $i,j,k\in \{x_0,x_1,x_2,x_3\}$, the superscripts denote contravariant components, while the subscripts denote covariant components. The spacetime in Einstein's general theory of relativity is a torsion-free manifold. Consequently, the affine connection $\Gamma_{ij}^k$ is symmetric and is entirely determined by the metric tensor as~\cite{KPThorne73book,Wald10book}
\begin{equation}\label{Eq: Gamma_GR}
\Gamma_{ij}^k = \frac{1}{2}g^{kl}(\partial_i g_{lj} + \partial_j g_{li} - \partial_l g_{ij}),
\end{equation}
known as the Christoffel symbol of the second kind, where $\partial_i$ represents $\partial/(\partial x_i)$, and $g_{lj}$ and $g^{kl}$ denote the covariant and contravariant metric tensor, respectively. 

Additionally, the Christoffel symbol of the first kind is defined by
\begin{equation}
\Gamma_{lij}=g_{lk}\Gamma_{ij}^k=\frac{1}{2}(\partial_i g_{lj}+\partial_j g_{li}-\partial_l g_{ij}),
\end{equation}
where all components are covariant. In quantum geometry, all quantities are typically treated as covariant. Since an additional band index $\nu$ is required, we will henceforth denote $\Gamma_\nu^{lij}$ [Eq.~\eqref{Eq: Christoffel}] as the quantum Christoffel symbol.

\renewcommand{\theequation}{B\arabic{equation}}
\setcounter{equation}{0}

\begin{figure}[tbp]
\centering 
\includegraphics[width=0.48\textwidth]{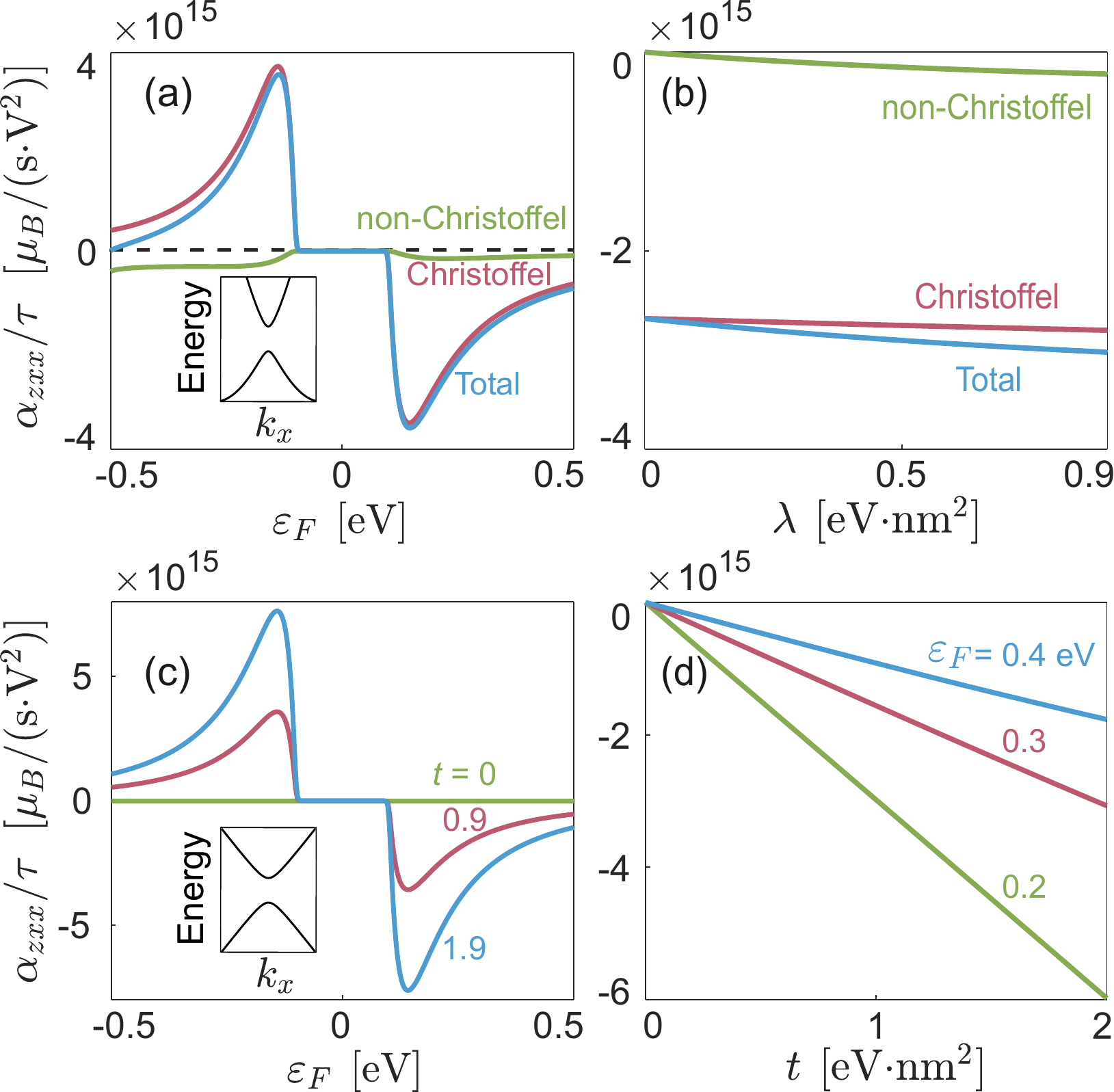}
\caption{Nonlinear magnetization coefficient $\alpha_{zxx}$ for the effective model [Eq.~\eqref{Eq: model}]. (a)  $\alpha_{zxx}$ as a function of the Fermi energy $\varepsilon_F$ without particle-hole symmetry ($\lambda = 0.5$ eV$\cdot$nm$^2$), for the contribution from the quantum Christoffel nonlinear magnetization and non-Christoffel contribution. 
(b) $\alpha_{zxx}$ as a function of the particle-hole symmetry breaking parameter $\lambda$ at $\varepsilon_F = 0.2$ eV and $t = 0.9$ eV$\cdot$nm$^2$.
(c) $\alpha_{zxx}$ as a function of $\varepsilon_F$ for the particle-hole symmetric case ($\lambda = 0$) at several values of $t$, which breaks the $\mathcal{C}_{x,y}^2$ rotational and $\mathcal{M}_{x,y}$ mirror reflection symmetries.
(d) $\alpha_{zxx}$ as a function of $t$ at $\lambda = 0$ and $\varepsilon_F = 0.2$, 0.3, and 0.4 eV.
Insets in (a) and (c) compare the band structures without and with particle-hole symmetry, respectively. Other model parameters are $v = 1$ eV$\cdot$nm and $m = 0.1$ eV.}
\label{Fig: model}
\end{figure}

\hypertarget{Appendix B}{\textit{\textcolor{blue}{Appendix B: Minimal model for quantum Christoffel nonlinear magnetization.}--}} 
To have an intuitive picture of when the quantum Christoffel nonlinear magnetization can emerge, we perform calculations of the nonlinear magnetization coefficient $\alpha_{zxx}$ for a four-band massive Dirac model~\cite{ZhangSC16np,Shen17book}.
\begin{equation}\label{Eq: model}
\mathcal{H} = v (k_x \tau_x\sigma_x + k_y \tau_x\sigma_y) + m \tau_z + t k_x k_y \tau_z+\lambda k^2,
\end{equation}
where $v$, $m$, $t$, and $\lambda$ are model parameters, and $\tau_i$'s and $\sigma_i$'s denote two sets of Pauli matrices. The term $tk_xk_y\tau_z$ is included to break $\mathcal{C}_{x,y}^2$ (two-fold rotations about the $x$ and $y$ axes) and $\mathcal{M}_{x,y}$ (mirror reflections with respect to the planes normal to the $x$ and $y$ directions) symmetries, which is necessary for a nonzero  $\alpha_{zxx}$, as illustrated in Table~\ref{Tab: Symm_pg}. Moreover, the last term $\lambda k^2$ is introduced to break the particle-hole symmetry. Insets in Figs. \ref{Fig: model}(a) and \ref{Fig: model}(c) compare the band structures without and with particle-hole symmetry, respectively. When $\lambda\neq 0$, the nonlinear magnetization coefficient is determined by Eq.~\eqref{Eq: alpha_full}. Fig.~\ref{Fig: model} (a) compares $\alpha_{zxx}$ from the quantum Christoffel nonlinear magnetization and non-Christoffel contribution in absence of particle-hole symmetry. Over the entire range of Fermi energy, the Christoffel contribution is dominant. As shown in Fig. \ref{Fig: model} (b), with increasing 
$\lambda$, that is, stronger violation of particle-hole symmetry, non-Christoffel contribution increases, although Christoffel contribution remains dominant.

For the particle-hole symmetric case ($\lambda=0$), the nonlinear magnetization is solely governed by the quantum Christoffel symbol $\Gamma_\nu^{ijk}$, and  $\alpha_{zxx}$ takes the form 
\begin{equation}\label{Eq: alpha_zxx}
\alpha_{zxx}=\frac{\tau e^3}{2\hbar\mathcal{V}}\sum_{\nu,\mathbf{k}}v_\nu^x f_0'(\Gamma_\nu^{xyx}-\Gamma_\nu^{yxx}).
\end{equation}
Fig.~\ref{Fig: model}(c) shows $\alpha_{zxx}$ as a function of the Fermi energy $\varepsilon_F$ for different values of $t$. The magnitude of $\alpha_{zxx}$ becomes pronounced near the band edge, where the quantum geometric contribution is significantly enhanced. Furthermore, as shown in Fig.~\ref{Fig: model}(d), a nonzero $t$ is essential for breaking the relevant symmetries and enabling a finite $\alpha_{zxx}$, consistent with the symmetry constraints in Table~\ref{Tab: Symm_pg}.

\hypertarget{Appendix C}{\textit{\textcolor{blue}{Appendix C: Details of First-principles calculations.}--}} 
The full (scalar)-relativity electronic structures of BiF$_3$ are calculated using density-functional theory within the pseudopotential framework. The projected augmented wave method, implemented in the Vienna \textit{ab initio} simulation package~\cite{kresse1996efficiency,kresse1999ultrasoft,kresse19962}, is employed for these calculations with the Perdew-Burke-Ernzerhof parameterization of the
generalized gradient approximation exchange-correlation functional~\cite{gga-pbe}. The electron wave function is expanded on a 6$\times$6$\times$1 $\mathbf{k}$-mesh through the self-consistent computation until the energy difference is less than 10$^{-6}$~eV. A cut-off energy of 400~eV is used for the plane-wave basis set.

To evaluate the quantum Christoffel nonlinear magnetization on a dense $\mathbf{k}$-mesh, we employ the Wannier interpolation scheme~\cite{wang-prb06} implemented in the WannierBerri code package~\cite{wannierberri}. The corresponding Wannier functions are constructed using the Wannier90 code package~\cite{wannier90}, starting from the atom-centered $s$ and $p$ orbitals of Bi and $p$ orbitals of F. The frozen window ranges from 0~eV to -6~eV referenced to the Fermi energy. The nonlinear magnetization coefficient, optical conductivity, and orbital magnetic moment of the unit cell are evaluated on a 600$\times$600$\times$1 $\mathbf{k}$-mesh using the symmetrized Wannier Hamiltonian, with 10 adopted refinement iterations~\cite{wannierberri}. 

In the simulation of nonlinear magnetization coefficient, the tetrahedron method~\cite{tetrahedronmethod,tetrahedronmethod2} is employed to accurately describe the distribution function $f_0'$ at zero temperature. In the simulation of optical conductivity, T = 100 K is used in the Fermi-Dirac distribution function $f_0$, as well as $\eta$ = 0.05 eV is chosen in Eq.~\eqref{Eq: opt-cond} to smooth the data curve.

\end{document}